# Use of Commutativity for Cryptology in Secret Communication

Mehmet Emir Koksal

*Department of Mathematics, Ondokuz Mayis University, 55139 Atakum, Samsun, Turkey*

emir_koksal@hotmail.com

**Abstract:** Commutativity of subsystems in cascade connected forms to form larger systems gets worthy to improve noise disturbance properties, stability, robustness and many other properties in system design. In this paper, another benefit of commutativity property is indicated, as far as the author knowledge for the first time, and illustrated by examples. This benefit is the gain of a new and original method for transmission of secret signals when travelling in a transmission channel. Hence, the paper presents an original and alternative method in cryptology. Their results are all validated by illustrative examples and Matlab simulation toolbox Simulink.

## I. Introduction

Second-order differential equations appears in many branches of engineering. They are used for a huge range of applications, including electrical systems, fluid systems, thermal systems and control systems. Especially, they are utilized as a powerful tool for modelling, analyzing, physical simulations and solving problems in modern and robust control theory, which is essential in any field of engineering and sciences.

In many cases, engineering systems are designed by interconnection of better simple first or second-order systems to achieve beneficial properties such as easy controllability, design flexibility, less sensitivity to disturbances and robustness. Feedback and cascade connection are among the commonly used interconnection structures in control and communication systems respectively. Cascade connection being an old but still an up to date designed method [1-4] can be used to improve further different system performances in connection with the commutativity concept. Commutativity of traditional linear time-invariant systems is straightforward; however linear time-varying systems have found many applications recently [5-10]. Therefore, the subject of this paper is devoted on the commutativity of linear time-varying systems only.

It is well-known that a cascade-connected systems is a combination of two subsystems so that the output of one is the input of the other [11]. If the input-output relation of the combination of two subsystems in cascade form is not effected by the order of the connection then, these two systems are set to be commutative [12].

There is a great deal of literature about the commutativity of continuous time-varying linear systems. Some of the important results about the commutativity are summarized in the sequel superficially.

J. E. Marshall is the first scientist studying on commutativity. In 1977, he proved that "for commutativity, either both systems are time-invariant or both systems are time-varying" [12]. Moreover, he proved necessary and sufficient conditions of first-order linear time-varying systems. Then, investigation of commutativity conditions for second-order, third-order and fourth-order continuous time-varying linear systems were studied in [13-15], [16] and [17] respectively together with some contributions appeared as conference presentations and a few short papers focusing to special cases such as first, second, third, and forth order systems.

In 1988, in [18], M. Koksal introduced the basic fundamentals of the subject [13], which is the first and one of two tutorial exhaustive journal papers. Another work joint by the same author has presented explicit commutativity conditions of fifth-order systems in addition to reviews of commutativity of systems with non-zero initial conditions, commutativity and system disturbance, commutativity of Euler systems [19].

In [20], all the second-order commutative pairs of a first-order linear time-varying analogue systems are derived. In [21], the decomposition of a second-order linear time-varying systems into its first-order commutative pairs are studied. This is important for the cascade realization of the second-order linear time-varying systems.

Even though there is a large cycle of works on the commutativity of continuous-time systems, there is only one journal literature on the commutativity of discrete-time systems [22].

Some benefits of commutativity of linear time-varying systems have already been appeared in the literature; for example designing systems less sensitive to parameter values [23], reducing noise interference and disturbance [20], improving robustness [19].

This paper focuses attention for a new encrypting method of obscuring the information transmitted through any communication channel by disguising it between transmitter and receiver. More precisely consider a communication system as shown in Fig. 1. In the figure, $A$ and $B$ represent commutative linear time-varying systems so that both channels $AB$ and $BA$ produce the same output signal for any applied input signal. But the transferred signal from transmitter to receiver proceeds in completely different shapes through the transmitting medium. Hence, this generates somewhat prevention against the infiltrators to stealing the secret information during transmission. More professional communication structure is indicated by using a single transmission channel which is used by time-sharing between two channels of Fig. 1 is shown in Fig. 2.

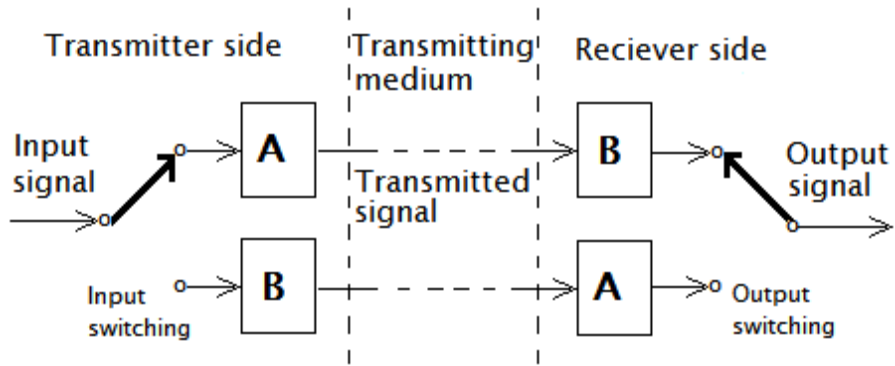

**Figure 1:** Transmitting a secret input signal in different transmitted forms through a double transmission channel.

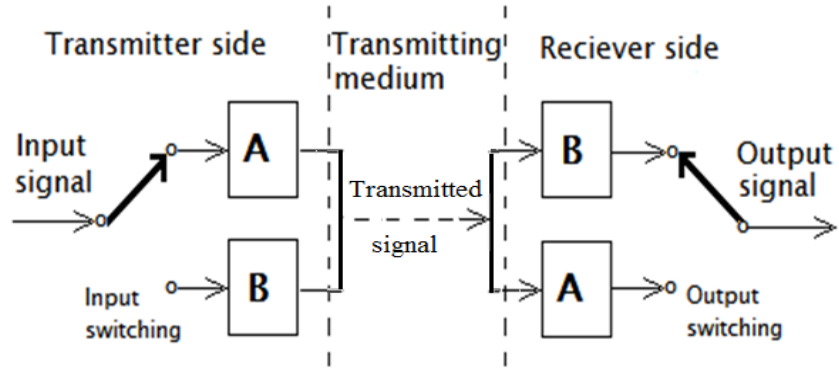

**Figure 2:** Transmitting a secret input signal in different forms through a single transmission channel by switching.

The above concept can be extended to more complex structures by using higher number of switching greater than 1. For example, with two identical subsystems $A$ and two identical subsystems $B$ (commutative with $A$) 4 communication passages of the input signal can be achieved through transmitting medium to obtain the same output signal. In fact, the structures $A \to ABB$, $AA \to BB$, $AAB \to B$, $AB \to AB$ where the arrow " $\to$ " separates the subsystems appearing in the transmitter and receiver sides. All these structures transfer any input signal to the same output signal which is transmitted in different shapes in transmitting medium by all of four structures. The concept can be extended for more complicated cases by using more than two different commutative pairs.

Since the concept of the use of commutativity property has already been introduced in Section 1, the paper is organized as to illustrate this original application by an example in Section 2. In the case of transmission using a single transmission channel which must be used time-sharing for transmission paths $A \to B$ and $B \to A$, switching is necessary; switching and switching effects are investigated in Section 3. Further applications is possible by using nearly commutative (or pseudo-commutative) subsystems and this is subjected in Section 4. For the same input-output signal pairs, the frequency spectrums of different transmitted signals proceeded in the transmission channel are compared in Section 5 to better illustrate the mentioned differences. Finally, the paper ends with Section 6 which includes conclusions.

## 2. Illustrative Example

To see how any input signal is transmitted to the same output signal in different forms of the transmitting medium, consider the communication structure in Fig. 1 with the following example:

**Example 1**

Let the linear time-varying subsystems A and B described by

$$A: \ddot{y}_A + (2 + 2sinw_0 t)\dot{y}_A + \left(5 - \frac{1}{2}cos2w_0 t + 2sinw_0 t + w_0 cosw_0 t\right) y_A = x_A, \quad (1a)$$

$$B: \frac{1}{2}\ddot{y}_B + \left(\frac{3}{4} + sinw_0 t\right)\dot{y}_B + \left(\frac{409}{32} - \frac{1}{4}cos2w_0 t + \frac{3}{4}sinw_0 t + \frac{1}{2}w_0 cosw_0 t\right) y_B = x_B, \quad (1b)$$

where $x_i$ and $y_i$ represent the input and output, respectively, of the subsystems $i = A, B$: (Double) dot on the top indicates (second) time derivative.

It is straight forward to show that $A$ and $B$ are commutative since the time-varying coefficients of $B$ can be obtained from those of $A$ by the relation (3a) in [20]

$$\begin{bmatrix} b_2(t) \\ b_1(t) \\ b_0(t) \end{bmatrix} = \begin{bmatrix} a_2(t) & 0 & 0 \\ a_1(t) & a_2^{0.5}(t) & 0 \\ a_0(t) & f_A(t) & 1 \end{bmatrix} \begin{bmatrix} k_2 \\ k_1 \\ k_0 \end{bmatrix}, \quad (2a)$$

where

$$k_2 = \tfrac{1}{2}, k_1 = -\tfrac{1}{4}, k_0 = \tfrac{4213}{400} \quad (2b)$$

and

$$f_A = \tfrac{2a_1 - \dot{a}_2}{4\sqrt{a_2}} = 1 + sinw_0 t. \quad (2c)$$

Since $k_1 \neq 0$, the second one of the sufficient conditions of commutativity

$$A_0(t) = a_0 - f_A^2 - \sqrt{a_2} f_A(t) = 3.5 \quad (3)$$

is satisfied since $A_0(t)$ in Eq. (3c) is constant (See Eq. (2.b). in [22]). It is easy to show that when the average values of coefficients are considered both systems are asymptotically stable with eigenvalues

$$A_{1,2} = -1 \pm j2, \quad (4a)$$

$$B_{1,2} = -\tfrac{3}{4} \pm j5. \quad (4b)$$

This implies though not guaranties, the high possibility of stability of actual time-varying subsystems $A$ and $B$ defined by Eqs. (1a) and (1b), respectively [24]; in fact, simulation results show that both systems are asymptotically stable.

To observe that both of the switching alternatives $A \to B$ and $B \to A$ shown in Figure 1 where $A$ and $B$ are defined in Eqs. (1a) and (1b) with $w_0 = 2\pi$ yield the same output at the receiver side, an input signal ($30 sin 1.2\pi t$ + a saw-tooth of period 3.3s and increasing from $-30$ to $+30$) is applied on the transmitter side. As observed in Figure 3, the transmissions $A \to B$ and $B \to A$ yield the same output signal (see ---- Output signal *10). In spite of the same input-output pairs for switchings $AB$ and $BA$, the travelled signals processed through transmission medium (see .... Transmitted signal $A - B$, .... Transmitted signal $B - A$) are quite different.

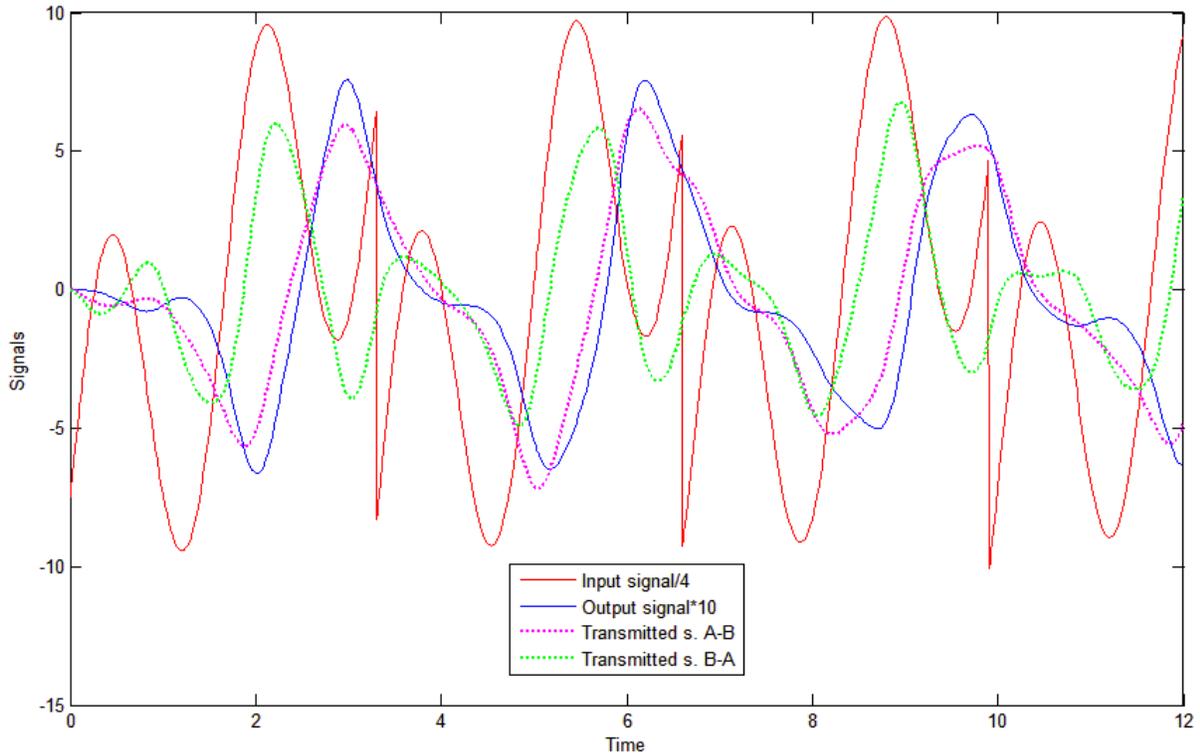

**Figure 3:** Input, output and transmitted signals in the communication system of Example 1.

To verify that the discussions are independent of the input signal applied, the simulations are repeated with a pulse train of amplitude 30, period 5 and with a pulse with of 10 %. The input signal and the same output of both transmission switching paths $A \to B$ and $B \to A$ are shown in Figure 4 (see —— Input signal/10, Outpu signal*10, respectively). It is also seen in this figure that the signals proceeded through the transmission medium, namely (....Transmitted s. $A - B$) and (.... Transmitted s. $B - A$), are quite different. Hence the same output signal is received by channels *AB* and *BA* for the same input signals irrespective of the shape of the input signal whilst different signals are transmitted through the transmission medium.

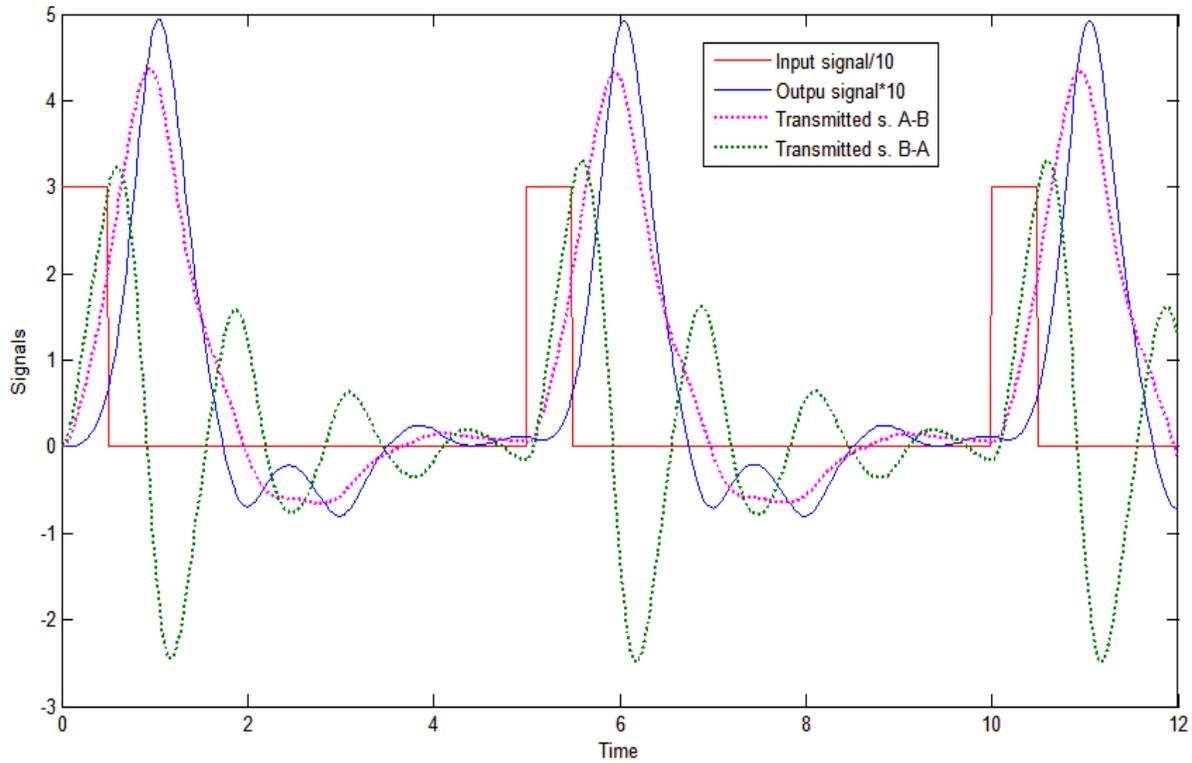

**Figure 4:** Input, output and transmitted signals in the communication system in Example 1 for a different input.

## 3.    Effects of Switching on Commutativity

When commutativity concept is used to transmit a signal through a transmission medium secretly using a single transmission channel time shared by transmissions $A \to B$ and $B \to A$ as described in Section 5.1, a sufficiently high rate of switching is necessary to puzzle malicious persons or infiltrator to resolve the transmitted information. At the beginning of each transmission slots (say $A \to B$) some initial conditions had been formed in the subsystems *A* and *B* during the previous slot ($B \to A$ in this case) and these initial conditions may not satisfy the second commutativity condition for unrelaxed systems $A$ and $B$ at the initial time of the current time slot [6 , Theorem 3.1 (Koksal 2)]. Therefore the output of the switched transmission system will be different from those of non switched systems $AB$ and $BA$. This difference will be more appreciated if the damping properties of subsystems $A$ and $B$ are weak (time constants are large with respect to switching period). This is because such properties will elongate natural responses of subsystems and the effect due initial conditions that had been formed improperly for commutativity in the previous time slot. Hence, the subsystems used in single channel transmitting system described in Section 5.1 better to have high damping coefficients for the output signal not effected by the switching considerably. This argument will be illustrated by the following two examples, namely Examples 2 and 3:

**Example 2**

To illustrate the above mentioned effect first consider the systems $A$ and $B$ which are commutative under zero initial conditions:

$$A: \dot{y}_A(t) + (1 + cos\pi t)y_A(t) = x_A(t); \quad y_A(0) = 0, \tag{1a}$$

$$B: \dot{y}_B(t) + (2 + cos\pi t)y_B(t) = x_B(t); \quad y_B(0) = 0. \tag{1b}$$

Where $B$ is obtained from $A$ by using constants $c_1 = c_0 = 1$(see Eq. 26 in [24]); namely,

$$\begin{bmatrix} b_1 \\ b_0 \end{bmatrix} = \begin{bmatrix} 1 & 0 \\ 1 + cos\pi t & 1 \end{bmatrix} \begin{bmatrix} c_1 \\ c_0 \end{bmatrix} = \begin{bmatrix} 1 & 0 \\ 1 + cos\pi t & 1 \end{bmatrix} \begin{bmatrix} 1 \\ 1 \end{bmatrix} = \begin{bmatrix} 1 \\ 2 + cos\pi t \end{bmatrix}. \tag{2}$$

Note the eigenvalues of $A$ and $B$ are $\lambda_A(t) = -1 - cos\pi t$ and $\lambda_B(t) = -2 - cos\pi t$, respectively. These eigenvalues remain in the left half of $s-$plane all the time (except the instants $t = 1.3.5, ...$ when it moves to origin instantly for subsystem $A$), hence both are likely asymptotically stable [25].

For the commutativity of subsystems under non zero initial conditions as well, it is required by the above mentioned second commutativity condition (Theorem 3.1 in [19], Eq. 27 in [21]) that

$$c_1 + c_0 = 1, \tag{3a}$$

$$y_A(t_s) = y_B(t_s), \tag{3b}$$

where $t_s$ is any switching instant (see Eq. 11 and 12 in [24], respectively). When either one or two of these conditions are not satisfied, the systems $AB$ and $BA$ may not have the same output when excited by any input. In the present example Condition in Eq. (3a) is not obviously satisfied since $c_1 + c_0 = 2 \neq 1$; considering the second condition, there is no guaranty that Eq. (3b) will be valid at the initial time of any switching slot since the initial conditions have been formed in the previous slot according to the dynamics of $A$ and $B$ rather independently. But, it is intuitively expected that the initial condition responses will decay fast to zero for highly damped subsystems and the outputs of $AB$ and $BA$ will dominantly be determined by the forced response generated by the input signal. Hence the coherence between the outputs $AB$ and $BA$ will not be effected considerably due to nonsatisfaction of the second commutativity condition spoiled by switching. The subsystems $A$ and $B$ defined by Eqs. (1a) and (1b), respectively, are examples of low damping systems (compared to subsystems that will be considered in Example 3), so that switching is expected to will cause a great difference in the outputs when compared with the same output of $AB$ and $BA$ resulted without switching.

For an input $10 sin 2\pi t + sawtooth\ wave\ with\ period\ 3\ magnitude \pm 30$ both systems $AB$ and $BA$ give the same output (—— Output signal*2.5) as shown in Fig. 5; on the same figure the input signal the transmitted signal $A \to B$ and $B \to A$ are shown by (—— Input signal/2), (····Transmitted s. $A - B$), (···· Transmitted s. $B - A$), respectively.

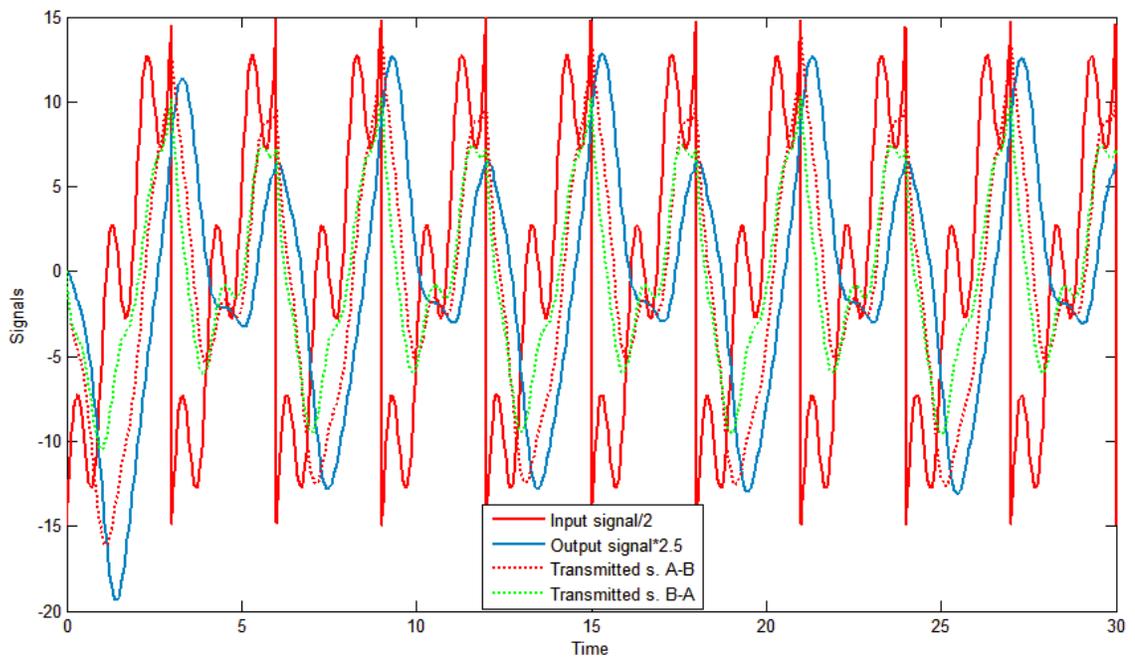

**Figure 5**. Input, Output and Transmitted signals by transmission paths $A \to B$ and $B \to A$ for Example 2

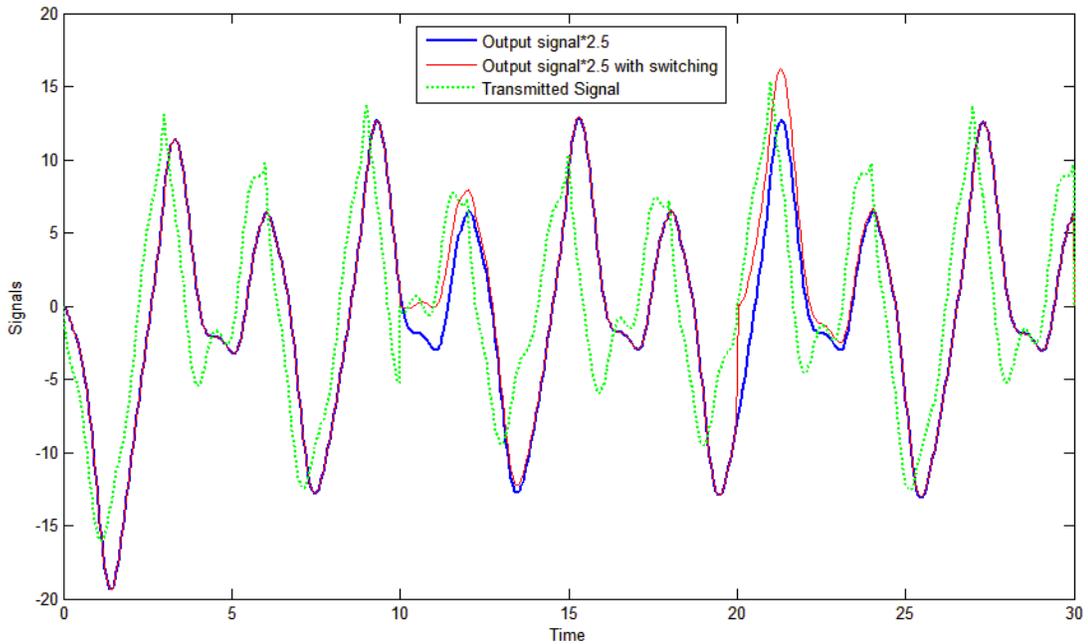

**Figure 6.** Transmitted signal in the single channel transmission and comparison of the output signals with and without switching for Example 2.

To observe the effect of switching on the shape of the output signal, the path $A \rightarrow B$ and $B \rightarrow A$ are switched periodically in sequence for durations of 10 seconds. The output signal at the receiver end is shown in Fig. 6 (—— Output signal *2.5 with switching); on the same figure, the output signal of connections $AB$ and $BA$ which appear in Fig. 5 (—— Output signal*2.5) is replotted. As it is expected there is a difference between the direct communication with two lines without switching and communication by switching with a single transmission line; this difference is really apparent just after each switching instant for about 4-5 second duration and then disappear in the rest of the switching period so that the output coincides with the ideal case of direct communication without switching. This vacancy of switching can be reduced by using subsystems having higher damping. Fig. 6 also includes the transmitted signal on the single line time shared by transmissions $A \to B$ and $B \to A$ (···· Transmitted signal).

**Example 3**

To observe the reduction of difference between two channel transmissions without switching and single channel transmission with switching, we consider similar subsystems in Example 2 but having relatively high damping then subsystems of Example 2.

Consider the systems $A$ and $B$ which are commutative under zero initial conditions:

$$A: \dot{y}_A(t) + (5 + cos\pi t)y_A(t) = x_A(t); \ y_A(0) = 0, \tag{4a}$$

$$B: \dot{y}_B(t) + (2 + cos\pi t)y_B(t) = x_B(t); \ y_B(0) = 0, \tag{4b}$$

where $B$ is obtained from $A$ by using constants $c_1 = 1, c_0 = -3$ (see Eq. 26 in [24]); namely,

$$\begin{bmatrix} b_1 \\ b_0 \end{bmatrix} = \begin{bmatrix} 1 & 0 \\ 5 + cos\pi t & 1 \end{bmatrix} \begin{bmatrix} c_1 \\ c_0 \end{bmatrix} = \begin{bmatrix} 1 & 0 \\ 5 + cos\pi t & 1 \end{bmatrix} \begin{bmatrix} 1 \\ -3 \end{bmatrix} = \begin{bmatrix} 1 \\ 2 + cos\pi t \end{bmatrix}. \tag{5}$$

Note the eigenvalues of $A$ and $B$ are $\lambda_A(t) = -5 - cos\pi t$ and $\lambda_B(t) = -2 - cos\pi t$, respectively. These eigenvalues remain in the left half of $s-$plane all the time; in fact, exception of some instants for subsystem $A$ in Example 2 does not occur in this case. Note also that the subsystem of Example 2 having larger damping or being more stable (subsystem $B$) is preserved in this example. Hence, speaking about both subsystems generally, those of Example 3 are more likely asymptotically stable and have higher damping than those of Example 2 [25]. Therefore, due to the reasons explained in Example 2, using a single channel by switching is expected to cause less deviation in the outputs when compared with the same output of $AB$ and $BA$ resulted without switching.

For the same input as in Example 2 both systems $AB$ and $BA$ give the same output (—Output signal*2.5) as shown in Fig. 7; on the same figure the input signal, the transmitted signal $A \to B$ and $B \to A$ are shown by (── Input signal/2), (···Transmitted s. $A - B$), (··· Transmitted s. $B - A$), respectively.

To observe the effect of switching on the shape of the output signal, the path $A \to B$ and $B \to A$ are switched periodically in sequence for durations of 10 seconds. The output signal at the receiver end is shown in Fig. 8 (── Output signal *2.5 with switching); on the same figure, the output signal of connections $AB$ and $BA$ which appear in Fig. 5 (── Output signal*2.5) is replotted. As it is expected there is a difference between the direct communication with two lines without switching and communication by switching with a single transmission line; this difference is really apparent just after each switching instant for about smaller than 1.5 second duration and then disappear in the rest of the switching period so that the output coincides with the ideal case of direct communication without switching. Note that this vacancy of switching is reduced from 5-7 seconds of Example 2 to values less than 1.5 by using subsystems having higher damping in

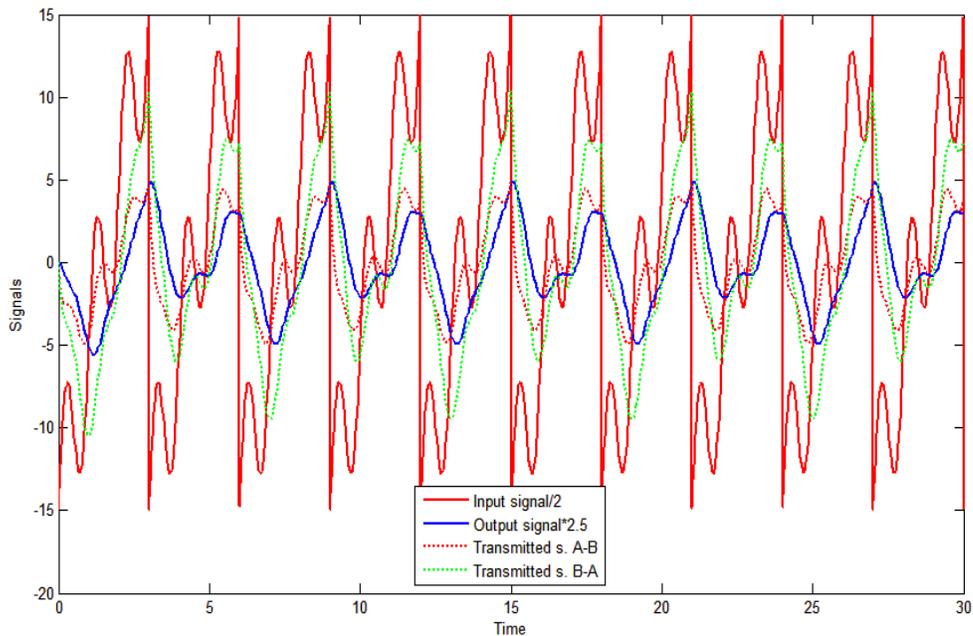

**Figure 7.** Input, Output and Transmitted signals by transmission paths $A \rightarrow B$ and $B \rightarrow A$ for Example 3

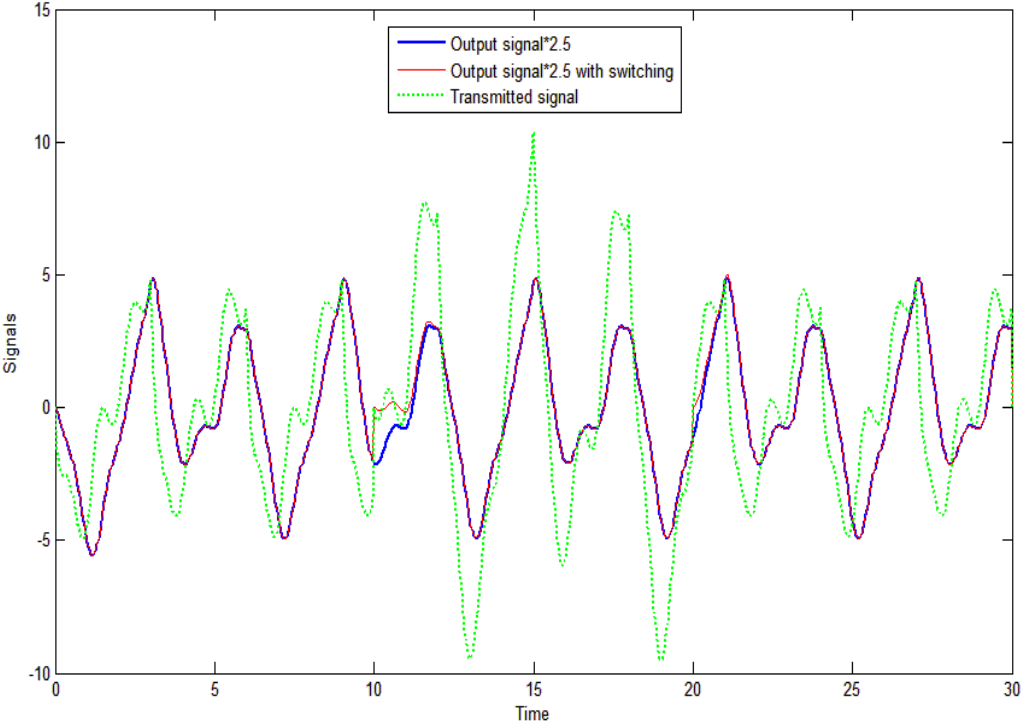

**Figure 8.** Transmitted signal in the single channel transmission and comparison of the output signals with and without switching for Example 2.

Example 3. Fig. 8 also includes the transmitted signal on the single line time shared by transmissions $A \rightarrow B$ and $B \rightarrow A$ (⋯ Transmitted signal). As a conclusion, effects of switching between the transmission channels can be better reduced by using highly damped subsystems $A$ and $B$.

4. **Use of Pseudo-Commutative Subsystems**

In the following example, how the use of commutativity in cryptology can be expanded by using the concept of nearly-commutative subsystems. Since the purpose is just to introduce this goal, first order subsystems are considered for simplicity.

**Example 4**

Let $A$ be the first order linear time-varying system of Example 3, namely defined by

$$A: \dot{y}_A(t) + (5 + cos\pi t)y_A(t) = x_A(t); \quad y_A(0) = 0. \tag{6a}$$

The system is chosen purposely as to have highly damped characteristic value always remaining in the left half of s-plane far away from the imaginary axis. This is because to have sufficiently fast decaying natural responses due to mismatching initial conditions preventing commutativity as mentioned before. Otherwise, when switching take place as described in Section 5.2 nonzero initial conditions not satisfying the commutativity conditions that had been formed before the switching from $A \to B$ to $B \to A$ or vice and these paths will not give the same responses. We now consider the transformation in Eq. (5) which gives all the commutative pairs of Subsystem $A$; instead of choosing $c_1$ and $c_2$ as constants, let us choose them as parameters varying slowly with respect to the natural dynamics of Subsystem $A$ and $B$. This is expected to result with nearly commutative subsystems so that the cascade connections $AB$ and $BA$ will yield almost the same outputs whilst the spectrum of the transmitted signal trough the communication medium continuously changing with varying parameters $c_1$ and $c_2$; this will puzzle the third persons trying to catch the actual information illegally. To observe this, we choose $c_1 = 2 + sin0.1\pi t$ and $c_0 = -3cos0.2\pi t$ in the mention respect and by a similar equation to Eq. (5) we obtain Subsystem $B$ as

$$B: (2 + sin0.1\pi t)\dot{y}_B(t) + (105sin0.1\pi t - 3cos0.2\pi t)y_B(t) = x_B(t); \quad y_B(0) = 0. \tag{6b}$$

For an input $x(t) = 12sin2\pi t$ which is shown in Fig. 9 (—Input/10), the outputs of transmissions $A \to B$ and $B \to A$ are also plotted in the figure (—Output*50: $AB$ and —Output*50: $BA$, respectively). Even though $A$ and $B$ are not exactly commutative, It is seen that $AB$ and $BA$ almost produce the same output owing to the slow variations of parameters $c_1$ and $c_0$ used to obtain $A$ from $B$ through a similar equation to Eq. (5). Further, on the same figure is shown the output (—Output*50: $AB - BA$ switched.) of the single line system which is used time-sharing between transmissions $A \to B$ and $B \to A$. It is obvious that switching used for a single channel transmission does not spoil to achieve the same response of systems $AB$ and $BA$.

In Fig. 10, it is seen that although all outputs are almost the same, the shape of the proceeding signals on the channel $A \to B$ (—Transmitted s. $AB$), on the channel $B \to A$ (—Transmitted s. $BA$), and on the common channel (...Transmitted s. $AB - BA$ switched) quite different. Hence, the same output signal is transferred through the transmission medium in different forms and this complicates attaining it by unauthorized persons.

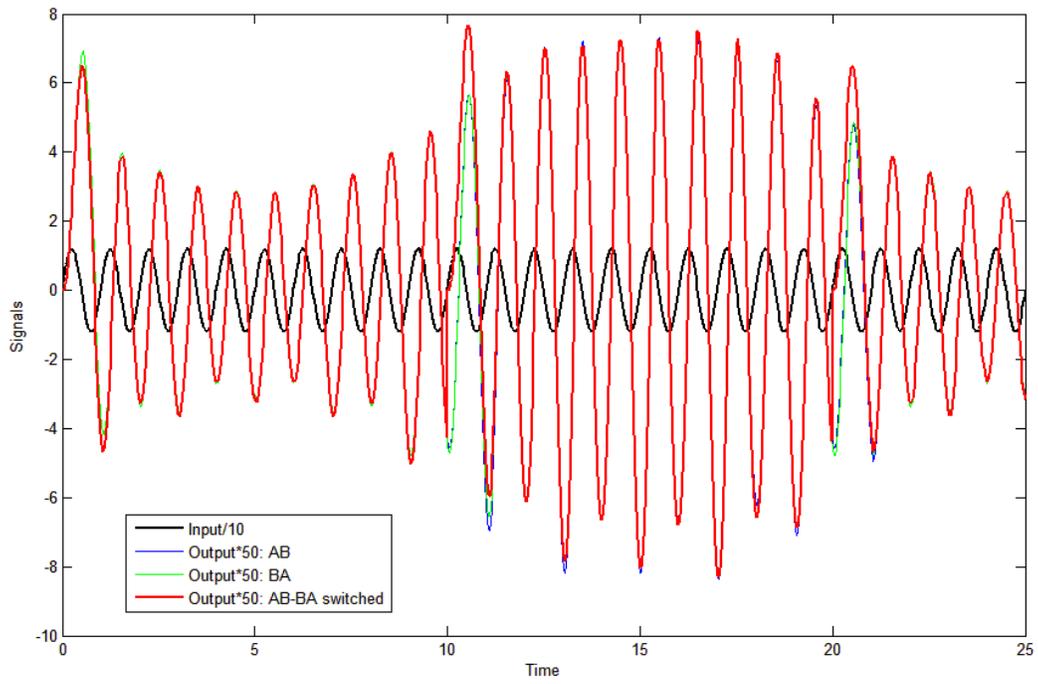

**Figure 9.** Output signals obtained at the receiver side for Example 4.

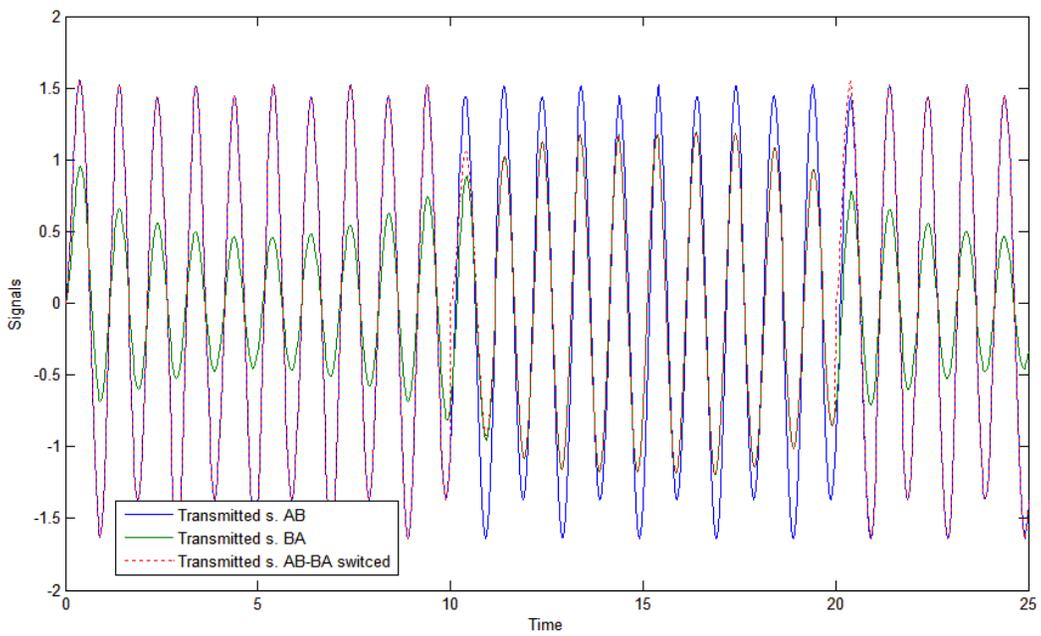

**Figure 10.** Transmitted signals travelling through transmission medium for Example 4.

## 5. Power Spectrums

On the base of their frequency spectrums, this section explains the use of commutativity for encrypting signals when travelling through the transmission channels. Since the power or frequency spectrum is an important subject for comparing signals and is a very essential analysis tool in communication theory, especially for studying different modulation techniques, the comparison of signals transmitted through the transmission medium depicted in Fig. 1 by comparing their spectrums is essential and this is considered in this section.

For Example 1, the spectrums of the transmitted signal from Subsystem $A$ to Subsystem $B$ and the transmitted signal from Subsystem $B$ to Subsystem $A$ are shown in Fig. 11 by (__Transmitted $AB$) and (__Transmitted $BA$), respectively. Obviously the spectrums are quite different in spite of the fact that these signals produce the same outputs at the receiver end (See ___Output signal*10 in Fig. 3).

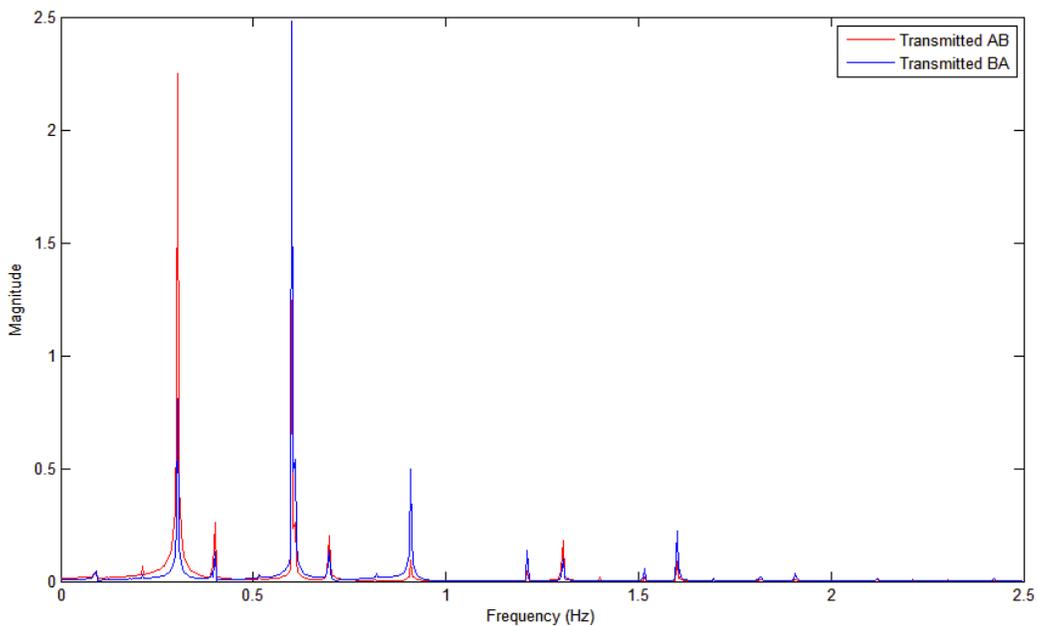

**Figure 11.** Power spectrums of the signals transmitted $A \rightarrow B$ and $B \rightarrow A$ for Example 1.

It has been already noted that the distortive effects of using switching for single channel communication as described in Section 3 and pseudo-commutative subsystems to strength cryptological actions as illustrated in Section 4 are hardly seen at the receiver output, the propagating signals from transmitter to receiver side depicted in Figs. 5, 7, 10 will naturally contain quite different spectrums similar to those in Fig. 11; whilst they are producing almost the same output signal on the receiver side (See Figs. 6, 8, 9, respectively). Therefor it is satisfied with this much dealing about the frequency spectrums characteristics.

## 6. Conclusions

Cryptology is an important subject for hiding signals in communication systems transferring information from one local area to another. In this paper, how commutativity property of subsystems in a communication system can be used for transmitting signals safely by reducing the probability of stealing by unauthorized persons. In fact, it is shown that the same output signal at the receiver side of a communication channel can be transmitted simultaneously through the same channel by using commutative subsystems at the transmitter and receiver sides together with switching while changing its transmitted version through the transmission medium.

Instead of fixing some system parameters as in $c_1$ and $c_0$ in Example 4, some certain time-change for $c_1$ and $c_0$, for example changing them arbitrarily but slowly with time, will yield using pseudo-commutative subsystems and thus additional alternatives for hiding the transmitted information.

Moreover, the transmitted signal trough the transmission medium in case of communication on a signal channel can be further puzzled by changing the switching strategy; for example, changing the switching frequency and switching periods of channels AB and BA unsymmetrically will produce extra advantages for a safe communication. And this can be forwarded as a further research subject on the area.

**Acknowledgments:** This study was supported by the Scientific and Technological Research Council of Turkey (TUBITAK) under the project no. 115E952.